\documentclass[useAMS,usenatbib]{mn2e}
\usepackage{graphicx}
\newcommand{\kms}{km~s$^{-1}$}
\newcommand{\Msol}{M$_\odot$}
\newcommand{\Htwo}{H$_2$}

\title[The CO content of LG dIrrs]{The CO content of the Local Group
  dwarf irregular galaxies IC5152, UGCA438, and the Phoenix dwarf}

\author[Buyle, Michielsen, De Rijcke, Ott,
  Dejonghe]{P. Buyle$^{1}$\thanks{E-mail: Pieter.Buyle@UGent.be,
  dolf.michielsen@nottingham.ac.uk, sven.derijcke@UGent.be,
  Juergen.Ott@csiro.au}\thanks{Post-doctoral Fellow of the Fund for
  Scientific Research - Flanders, Belgium (F.W.O.)},
  D. Michielsen$^2$, S. De Rijcke$^1$\thanks{Post-doctoral Fellow of
  the Fund for Scientific Research - Flanders, Belgium (F.W.O.)},
  J\"urgen Ott$^3$\thanks{Bolton Fellow}, and H. Dejonghe$^1$\\ $^{1}$
  Sterrenkundig Observatorium, Ghent University, Krijgslaan 281, S9,
  B-9000 Ghent, Belgium\\ $^{2}$ School of Physics and Astronomy, University of
  Nottingham, University Park, Nottingham NG7 2RD, UK\\ $^{3}$ CSIRO
  Australia Telescope National Facility Cnr Vimiera \& Pembroke Roads
  Marsfield NSW 2122, Australia}
\begin{document}

\date{}

\pagerange{\pageref{firstpage}--\pageref{lastpage}} \pubyear{2006}

\maketitle

\label{firstpage}

\begin{abstract}
We present a search for CO($1 \rightarrow 0$) emission in three Local
Group dwarf irregular galaxies: IC5152, the Phoenix dwarf, and
UGCA438, using the ATNF Mopra radio telescope. Our scans largely cover
the optical extent of the galaxies and the stripped HI cloud West of
the Phoenix dwarf. Apart from a tentative but non-significant emission
peak at one position in the Phoenix dwarf, no significant emission was
detected in the CO spectra of these galaxies. For a velocity width of
6\,\kms, we derive 4$\sigma$ upper limits of 0.03\,K\,\kms,
0.04\,K\,\kms\ and 0.06\,K\,\kms\ for IC5152, the Phoenix dwarf and
UGCA438, respectively. This is an improvement of over a factor of 10
compared with previous observations of IC5152; the other two galaxies
had not yet been observed at millimeter wavelengths. Assuming a
Galactic CO-to-\Htwo\ conversion factor, we derive upper limits on the
molecular gas mass of $6.2\times 10^4$\,\Msol, $3.7\times
10^3$\,\Msol\ and $1.4\times 10^5$\,\Msol\ for IC5152, the Phoenix
dwarf and UGCA438, respectively. We investigate two possible causes
for the lack of CO emission in these galaxies. On the one hand, there
may be a genuine lack of molecular gas in these systems, in spite of
the presence of large amounts of neutral gas. However, in the case of
IC5152 which is actively forming stars, molecular gas is at least
expected to be present in the star forming regions. On the other hand,
there may be a large increase in the CO-to-\Htwo\ conversion factor in
very low-metallicity dwarfs ($-2 \le [$Fe/H$] \le -1$), making CO a
poor tracer of the molecular gas content in dwarf galaxies.
\end{abstract}


\section{Introduction}

Dwarf galaxies form the most abundant galaxy population in the nearby
universe.  They represent a wide class of low-mass, low-metallicity
stellar systems, with some being gas-rich and presently actively
forming stars (dwarf irregulars or dIrrs), while others are gas-poor
and apparently formed their stars long ago (dwarf spheroidals or
dSph). In between the above extremes exists a subclass of gas-rich
dwarfs, called transition-type galaxies, that are not currently
for\-ming stars. These objects occupy an evolutionary state
intermediate between dIrrs and dSphs \citep{mateo1998,grebel2001}.
%

Most studies of the interstellar medium (ISM) of dwarf galaxies up to
now have focused on their atomic gas (HI) content
\citep{young2003,conselice03,buyle05,bouchard05,tarchi05}. Nevertheless
if, in accordance with massive galaxies (e.g. the Milky Way), the star
formation predominantly takes place in molecular clouds, that are
observed to form within cold HI regions (with velocity dispersions of
the order of 3\,\kms), one would expect a correlation between the star
formation rate and the CO emission rather than between the star
formation rate and the amount of cold HI gas. However, so far no such
relation has been observed for dIrrs, neither for CO or HI. Generally
individual HI clouds are found close to star formation sites, with
usually an offset by a few 100\,pc. Also, some very dense HI clouds,
which surely exceed the density threshold above which star formation
is expected to be possible, show no signs of ongoing or recent star
formation \citep{mateo1998,begum06}. These results show that the
presence of a cold HI component may be a necessary ingredient for star
formation, but not a sufficient one: it seems that sometimes a trigger
is needed to initiate star formation in dense gas clouds
\citep{taylor1994,hernquist1995,taylor1997,pisano1999}. Moreover, the
LG (Local Group) dwarf galaxies show a very wide variety of gas masses
and star formation rates. Some are quite gas-rich but have little
ongoing star formation while others are gas-poor but show evidence for
recent star formation \citep{mateo1998}. In those LG dIrrs that have
been detected in CO, the emission usually comes from individual
clouds, about 50~pc wide \citep[see][and references
therein]{israel1997,mateo1998,derijcke06}. The locations of star
formation sites, traced by HII regions, correlate very well with
regions of CO emission. This shows that the emission of the molecular
ISM is a much better tracer of star formation than that of the cold
atomic ISM. In order to investigate this interplay between the
properties of the atomic and molecular ISM and star formation, we
observed a sample of nearby dIrrs:~IC5152, the Phoenix dwarf and
UGCA438. These galaxies have published HI masses but, with the
exception of IC5152, have not yet been surveyed for the presence of a
molecular ISM.

CO is often used to trace the molecular gas in galaxies because it is
the most abundant (after H$_2$) and brightest molecule. It is related
to the FIR luminosity of the galaxies that is emitted by the warm dust
heated by embedded high-mass OB stars, indicating that it indirectly
correlates with the current star formation rate
\citep[e.g.][]{gay2004,murgia2005}. It has also been found that dwarf
galaxies below a metallicity of $12 + \log({\rm O/H}) \sim 7.9$ are
not detected in CO emission \citep{taylor1998}. This suggests that the
CO to H$_2$ conversion factor ($X_{\rm CO}$) depends on metallicity
and that the CO-H$_2$ relation becomes sharply non-linear around this
metallicity \citep{taylor1998,taylor2001}. However, In studying these
low-metallicity dIrrs we might be able to shed new light on this
cut-off in CO detections.

In the next section we present more details on the three galaxies in our
sample. Section~\ref{sec_obs} describes the new CO observations, and the
results are presented and discussed in sections~\ref{sec_results} and
\ref{sec_disc} with a comparison with recent HI observations. Finally we
summarise our conclusions in section~\ref{sec_concl}.

\section{The sample}
\label{sec_sample}


\begin{table*}
\begin{center}
\begin{minipage}{140mm}
  \caption{Main galaxy properties. The radial velocities are taken from LEDA,
    the distances from \citep{karachentsev2004}.}
  \begin{tabular}{lccccccc}
    \hline
    Galaxy & $\alpha$ (J200) & $\delta$ (J2000) & $v_\odot$ & $D$ & $M_B$ & [Fe/H] & $12 + \log({\rm O/H})$\\
           & (h:m:s) & $(\deg:\arcmin:\arcsec)$ &   (\kms)  &(Mpc)& (mag) &  (dex) & (dex)\\
    \hline
    IC5152  & 22:02:42.5 & -51:17:47 &  $126 \pm 5$ & 2.07 & -17.2  & -1.0 & 8.00\\
    Phoenix & 01:51:06.3 & -44:26:41 & \ $56 \pm 6$ & 0.44 & \ -9.5 & -1.8 & 6.87\\
    UGCA438 & 23:26:27.6 & -32:23:20 & \ $62 \pm 4$  & 2.23 & -12.6  & -2.0 & 7.32\\
    \hline
  \end{tabular}
  \label{tab_properties}
\end{minipage}
\end{center}
\end{table*}

We selected a sample of 3 dwarf irregular galaxies which fulfilled our
selection criteria: being visible during night time, never targeted
before for CO observations, are nearby hence in the Local Group and
have published or archived (ATCA = Australian Telescope Compact Array)
HI observations \citep{stgermain1999,ott07}.

\subsection{IC5152}

IC5152 is classified as a dIrr at the outskirts of the LG, at a
distance of 2.07\,Mpc \citep{karachentsev2004}, and with a metallicity
[Fe/H$]\approx -1$ \citep{zijlstra1999}. Its properties are summarised
in Table~\ref{tab_properties}. The optical diameter of IC5152 is about
5\arcmin. The whole central region is a very active star formation
site, and several HII regions have been studied
\citep{webster1983,hidalgo2002,lee2003}. IC5152 contains $1.1 \times
10^8\,M_\odot$ of HI (Table~\ref{tab_results}) and is the most
gas-rich galaxy in our sample. The HI emission is centered on the
optical body of the galaxy, both projected on the sky
(Figure~\ref{fig_HImapIC}) as in velocity (Tables~\ref{tab_properties}
and \ref{tab_results}). 

\subsection{Phoenix dwarf}

The southern Phoenix dwarf galaxy, situated at a distance of 440~kpc
\citep{karachentsev2004}, is a very interesting system. It has a
heliocentric optical radial velocity of $56 \pm 6$\,\kms\ whereas the
HI clouds observed in the Phoenix vicinity have a heliocentric
velocity of $-23$\,\kms (Figure~\ref{fig_HImapPhoenix} and
Table~\ref{tab_results}), $59$\, \kms, $7$\, \kms\ and $140$\,
\kms. The HI clouds with a velocity of $7$\, \kms\ and $140$\, \kms\
are respectively associated with the Milky Way and the Magellanic
Stream. \citet{stgermain1999} found that the southern HI cloud with
velocity component $59$\, \kms\ is, because of its mass and large
offset ($\ge$1.2 kpc south) from the optical body of the galaxy, not
associated with Phoenix but with a HVC instead. The characteristics of
the HI clouds with velocity $-23$\,\kms\ however are consistent with
this gas having been associated with Phoenix in the past and being
lost by the galaxy after the last event of star formation in the
galaxy, about 100\,Myr ago \citep{gallart2001}. Possibly, the HI gas
was stripped by the ram pressure of the intergalactic medium
\citep{mayer2005}. If present, molecular clouds, being much more
compact than the atomic ISM, will not be stripped and could still be
present inside the galaxy's main body. It is a very metal-poor galaxy,
with [Fe/H$]=-1.8$ \citep{held1999}, with a morphology intermediate
between dIrr and dSph. The Phoenix dwarf contains $2.4\times
10^5$\,\Msol\ of HI (see Table~\ref{tab_results}).

\subsection{UGCA438}

\begin{figure}
  \includegraphics[scale=0.65]{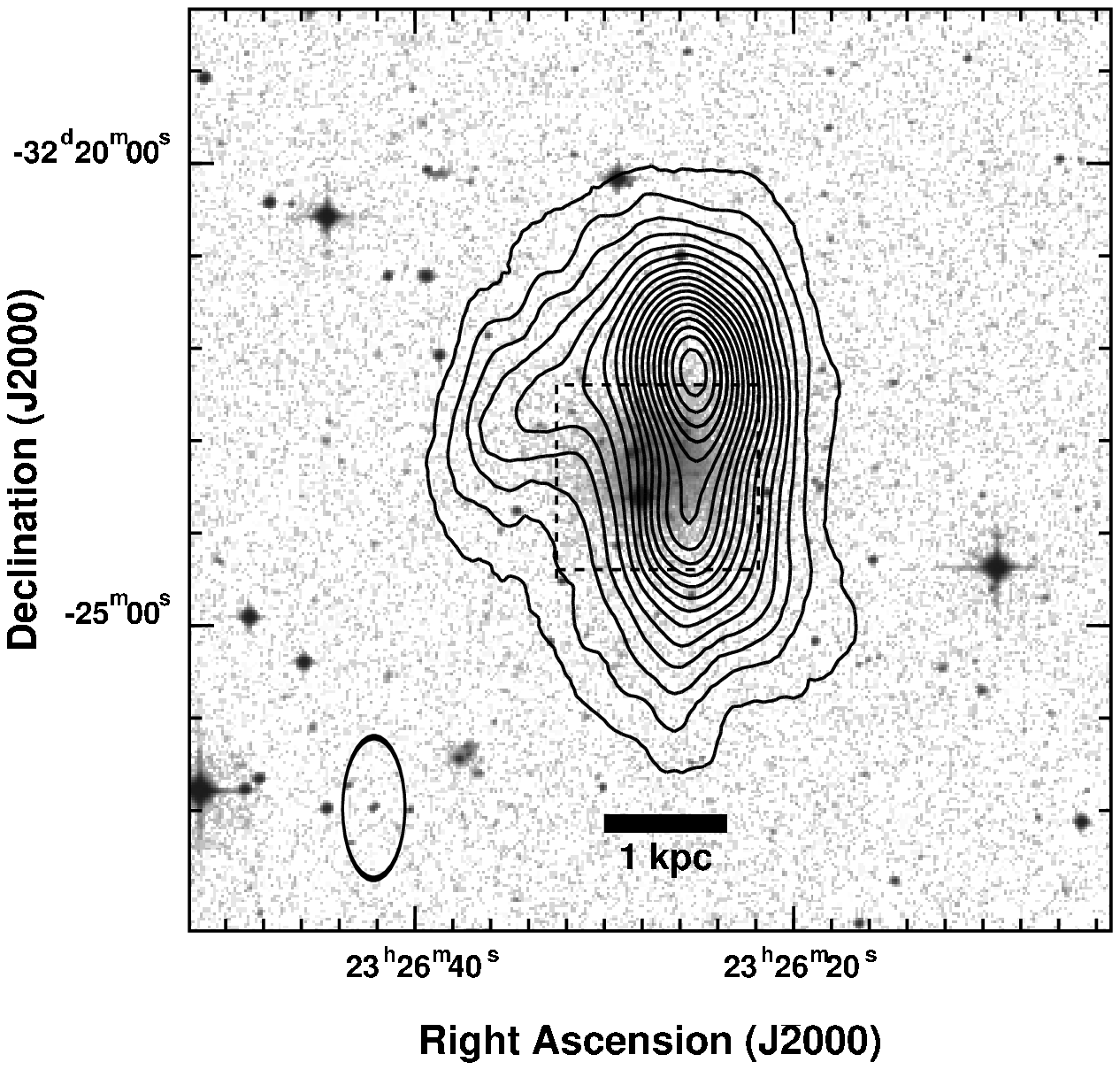} \caption{UGCA438: HI
  emission intensity contours (the outer contour indicates the
  3$\sigma$ $\sim 1.5\times10^{19}$\,cm$^{-2}$ level; the other
  contours each mark a 3$\sigma$ increase in intensity $\sim 3.0, 4.5,
  6.0, \ldots\times10^{19}$\,cm$^{-2}$) overlaid on an optical DSS
  image. The extent of the CO map is shown by the dashed
  rectangle. The ATCA beam of the HI observations is plotted in the
  bottom left corner. Original HI data from \citet{ott07}.}
  \label{fig_HImapUGCA}
\end{figure}

UGCA438 is a dIrr at a distance of 2.23\,Mpc, which places it just
outside the LG, making it a likely member of the Sculptor Group
\citep{lee1999,karachentsev2004}. With a metallicity [Fe/H$]=-2$, it
is one of the most metal deficient nearby dIrrs.  It contains a very
young population of stars, but no HII regions. UGCA438 has an HI mass
of $5.9 \times 10^6\,M_\odot$ (see Table~\ref{tab_results}). The HI
emission, although at the same radial velocity as and covering the
whole optical body of UGCA438, is slightly offset to the North of the
galaxy (Figure~\ref{fig_HImapUGCA}).

\section{Observations and data reduction}
\label{sec_obs}

The three galaxies were observed with the new 3mm receiver at the
Mopra\footnote{The Mopra radio telescope is part of the Australia
Telescope which is funded by the Commonwealth of Australia for
operation as a National Facility managed by CSIRO.} 22m single-dish
radio telescope, at Coonabarabran, Australia. The observations were
carried out during the first half of the nights between 2005 October
18 and 30, shortly after the Mopra upgrade. At the time of observing
the new spectrometer MOPS contained just one setup which produces 4097
channels across a 137.5 MHz band, which we centred on the CO($1
\rightarrow 0$) frequency (115.27 GHz), corrected for the Doppler
shift for each of the galaxies. The beamsize of the telescope at this
frequency is $\sim 33\arcsec$ and the velocity resolution of our
spectra is 0.08\, \kms.  The system temperature during the
observations varied between 450 and 600 K over the different nights,
depending on wheather conditions. The pointing of the telescope was
recalibrated every 1 to 1.5 hours by observing a nearby SiO maser
source (X Pav for IC5152, R Hor for Phoenix and R Aqr for UGCA438).

\begin{figure}
  \includegraphics[scale=0.65]{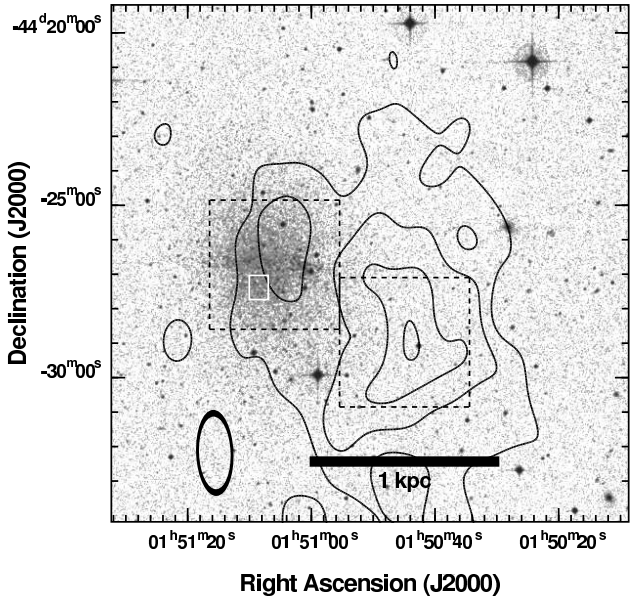} \caption{Phoenix: HI
  emission intensity contours (the outer contour indicates the
  3$\sigma$ $\sim2.1\times10^{19}$\,cm$^{-2}$ level ; the other
  contours each mark a 3$\sigma$ increase in intensity $\sim4.2, 6.3,
  8.4, \ldots\times10^{19}$\,cm$^{-2}$) overlaid on an optical DSS
  image. The extent of the CO maps is shown by the dashed
  rectangles. The white rectangle indicates the position of the
  possible CO detection (see Figure~\ref{fig_plotvel} and text for
  details). The ATCA beam of the HI observations is plotted in the
  bottom left corner. Original HI data from \citet{ott07}. }
  \label{fig_HImapPhoenix}
\end{figure}

Due to the proximity of the galaxies, their angular sizes are quite
large, hence mosaicing is the only way to cover their optical
bodies. We observed a field of $4.75\arcmin \times 4.75\arcmin$ for
IC5152, $1.75\arcmin \times 1.75\arcmin$ for UGCA438 and $4\arcmin
\times 4\arcmin$ for the Phoenix dwarf covering the entire optical
field of each galaxy. An additional field of $4\arcmin \times
4\arcmin$ which covers the stripped HI cloud was also observed, see
Fig.~\ref{fig_HImapPhoenix}.  We mapped the fields several times with
an 'on-the-fly' mapping procedure on a rectangular Nyquist grid.  The
mapping is done along equatorial coordinates, with each scan
perpendicular to the other to optimise the field coverage. Due to the
large size of the field, each scanned row was finalised with observing
an empty patch of sky next to the galaxy to accurately perform the sky
subtraction. This observing strategy is very different from the CO
observations of \citet{taylor1998} and \citet{taylor2001}, who target
just a few positions per galaxy, guided by the presence of cold H{\sc
i}.

\begin{figure}
  \includegraphics[scale=0.65]{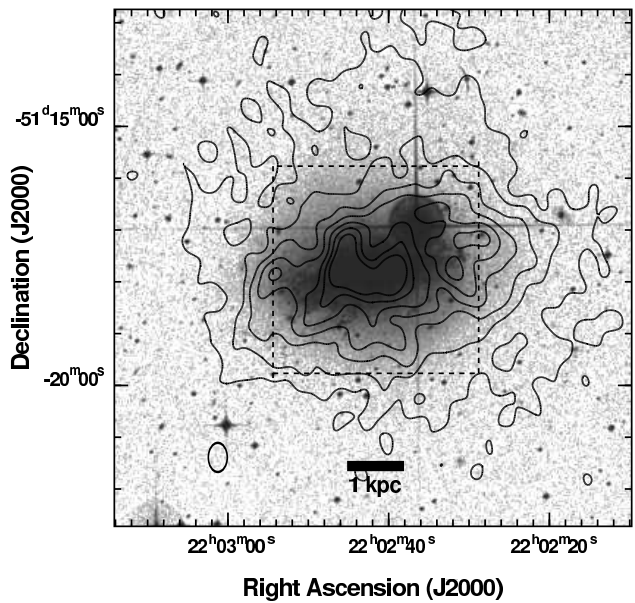} \caption{IC5152: HI
  emission intensity contours (the outer contour indicates the
  3$\sigma$ $\sim 2.7\times10^{20}$\,cm$^{-2}$ level; the other
  contours each mark a 3$\sigma$ increase in intensity $\sim 5.2, 7.9,
  10.6, \ldots \times10^{20}$\,cm$^{-2}$) overlaid on an optical DSS
  image. The extent of the CO map is shown by the dashed
  rectangle. The ATCA beam of the HI observations is plotted in the
  bottom left corner. Original HI data from \citet{ott07}.}
  \label{fig_HImapIC}
\end{figure}

The data reduction consists of three parts. First a correction has
been performed to align positional and time stamp information.
Afterwards the data is fed into Livedata, which performs a baseline
subtraction of the spectra. A polynomial of 2nd order was used for the
baseline subtraction. Finally, the spectra are spatially combined into
a data cube by means of the Gridzilla software.\footnote{Both Livedata
and Gridzilla are part of the larger AIPS++ package}. To convert the
CO flux from Kelvin to Jy, we use the Mopra receiver efficiency in the
CO band\footnote{This can be found in the Mopra guide
\texttt{http://www.narrabri.atnf.csiro.au/mopra/mopragu.pdf}}, which
is $\sim 30$\,Jy\,K$^{-1}$. We investigated the spectra unsmoothed and
smoothed to a velocity resolution of 1.28\, \kms. In both cases no
emission was detected. In order to derive upperlimits (see Table.~2),
we Hanning smoothed the spectra to a velocity resolution of $\pm$6\, \kms.

\subsection{HI observations}
All three galaxies, IC\,5152, UGCA\,438, and the Phoenix dwarf were
observed with the Australia Telescope Compact Array (ATCA). For the
Phoenix dwarf galaxy we re-reduced archival data (project C569, PI
Oosterloo, observed in 1996 September) which are originally published
in \citet{stgermain1999}. The beam and the 1$\sigma$ column density
sensitivity of the data, observed in the compact EW\,367 array
configuration, are $140\arcsec\times80\arcsec$ and
$\sim7\times10^{18}$\,cm$^{-2}$, respectively.

IC\,5152 was observed in three array configurations in 2000 January
(two 12\,h tracks in the 1.5 array configuration; project code C809,
PI de Blok), 2004 February (one track in the 750 configuration; C1253,
PI Ott), and 2004 April (one track in EW\,367; C1253). After standard
calibration and imaging, including CLEANing, the final data cube has a
beam of $35\arcsec\times22\arcsec$ with a 1$\sigma$ rms noise of
3.7\,mJy\,beam$^{-1}$ in a 3.3\,km\,s$^{-1}$ plane, resulting in a
1$\sigma$ rms column density sensitivity of $\sim
9\times10^{19}$\,cm$^{-2}$.

Similar to IC\,5152, UGCA\,438 was observed using three array
configurations, too. All data were collected within project C1253. The
three 12\,h tracks were conducted on 2003 November (1.5 array), 2004
February (750), and 2004 April (EW\,367). The resulting reduced data,
which will be presented in detail elsewhere, have a 1$\sigma$ rms
sensitivity of $\sim 3$\,mJy\,beam$^{-1}$ in a 1\,km\,s$^{-1}$ plane
with a beam of $84\arcsec\times46\arcsec$ in size. The resulting
1$\sigma$ column density sensitivity is $\sim
5\times10^{18}$\,cm$^{-2}$.

\section{Results}
\label{sec_results}

\begin{table*}
\begin{center}
\begin{minipage}{140mm}
  \caption{Properties of the cold and molecular ISM. The HI results
    (of all galaxies) are from \citet{ott07} and the CO results are
    4$\sigma$ upper limits, assuming a velocity width of 6\,\kms.}
  \begin{tabular}{lcccccc}
    \hline
   Galaxy & $M({\rm HI})$ & $v_{\rm HI}$ & $W_{50,\rm HI}$ & $S({\rm CO})$ & $M_{\rm mol}$\footnote{using Galactic $X_{\rm CO,Gal}$} & $M_{\rm mol}$\footnote{using metallicity dependent $X_{\rm CO,Z}$ of eqn. (\ref{XCO})} \\
          & (\Msol)       & (\kms)       & (\kms)             & (K \kms)           & (\Msol)           & (\Msol)  \\  \hline
    IC5152  & $1.1\times 10^8$ & $122\pm 5$ & $86\pm 5$ & $< 0.03$ & $< 6.2\times 10^4$ & $< 2.4\times 10^5$ \\
    Phoenix & $2.4\times 10^5$ & $-23\pm 3$ & $22\pm 3$ & $< 0.04$ & $< 3.7\times 10^3$ & $< 8.3\times 10^4$ \\
    UGCA438 & $5.9\times 10^6$ & $~19\pm 3$ & $19\pm 3$ & $< 0.06$ & $< 1.4\times 10^5$ & $< 1.6\times 10^6$ \\
    \hline
  \end{tabular}
  \label{tab_results}
\end{minipage}
\end{center}
\end{table*}

\begin{figure}
  \includegraphics[scale=0.6]{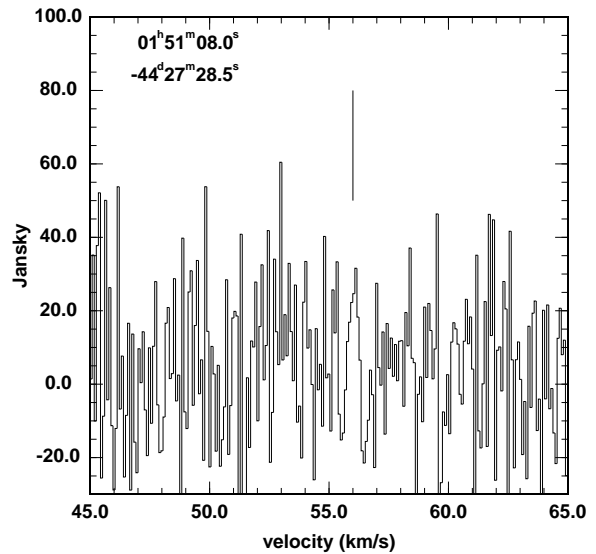}
  \caption{Spectrum of the Phoenix dwarf in the region indicated by a
    white rectangle in Figure~\ref{fig_HImapPhoenix}. At the optical
    velocity of Phoenix, $v_\odot = 56$\,\kms\, there is a hint of CO
    emission, in the form of 8 consecutive non-negative channels.}
  \label{fig_plotvel}
\end{figure}

In Figures~\ref{fig_HImapIC} to \ref{fig_HImapUGCA}, the dashed
rectangles indicate the area of the observed CO maps, covering the
optical extent of the galaxies. Although IC5152 and UGCA438 contain a
significant amount of neutral gas, we find no CO emission in these
galaxies. Also in the HI cloud of the Phoenix dwarf we find no
evidence for the presence of CO. Within the Phoenix dwarf itself, some
CO might be left. In Figure~\ref{fig_plotvel}, we show the spectrum of
the region indicated by a white rectangular in
Figure~\ref{fig_HImapPhoenix}. Exactly at the optical velocity of
Phoenix (56\,\kms), there is a small emission peak. If this feature is
real, it is unlikely to be related to galactic CO emission since
Phoenix lies about 70 degrees above the Galactic plane, which would
mean a molecular cloud moving almost perpendicular to the Galactic
plane with a velocity of 56\,\kms.  However, deeper observations are
needed to confirm the detection.

We provide two estimates for the upper limits on the molecular gas
masses of the observed galaxies, using on the one hand the Galactic
$X_{\rm CO,Gal} = 3 \times 10^{20}$\, cm$^{-2}$\, (K \kms)$^{-1}$
\citep{strong88,scoville87} and the metallicity dependent conversion
factor $X_{\rm CO,Z}$ given by
\begin{equation}
\log\left(\frac{X_{\rm CO,Z}}{X_{\rm CO,Gal}}\right) = (5.95 \pm 0.86) -
(0.67 \pm 0.10)\left(12+\log\left(\frac{O}{H}\right)\right) \label{XCO}
\end{equation}
\citep{roberts1991,wilson1995}. The CO flux and molecular gas mass
upper limits are listed in Table~\ref{tab_results}.

\section{Discussion}
\label{sec_disc}

\subsection{CO and metal abundance}

\citet{taylor1998} found that dwarf galaxies below a metallicity of
$12 + \log({\rm O/H}) \sim 7.9$, or $Z \leq 0.1 Z_\odot$ are not
detected in CO emission. Based on a study of 121 Northern
IRAS-detected dwarf galaxies, \citet{leroy2005} also find that
detections of galaxies with metallicities of $12+\log({\rm O/H}) \leq
8.0$ are quite rare.

\begin{figure*}
\vspace*{6cm}
\special{hscale=66 vscale=66 hsize=500 vsize=500
hoffset=-5 voffset=-235 angle=0 psfile="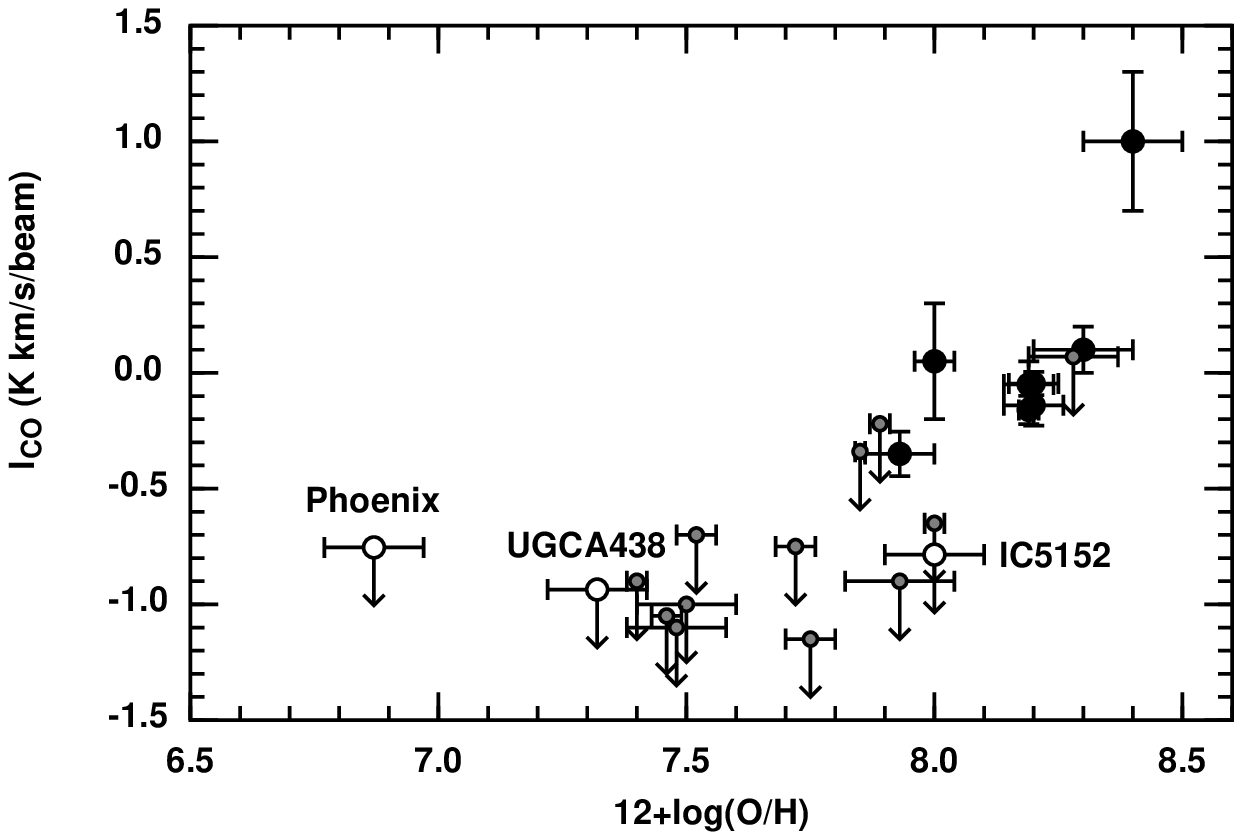"}
\special{hscale=66 vscale=66 hsize=500 vsize=500
hoffset=240 voffset=-235 angle=0 psfile="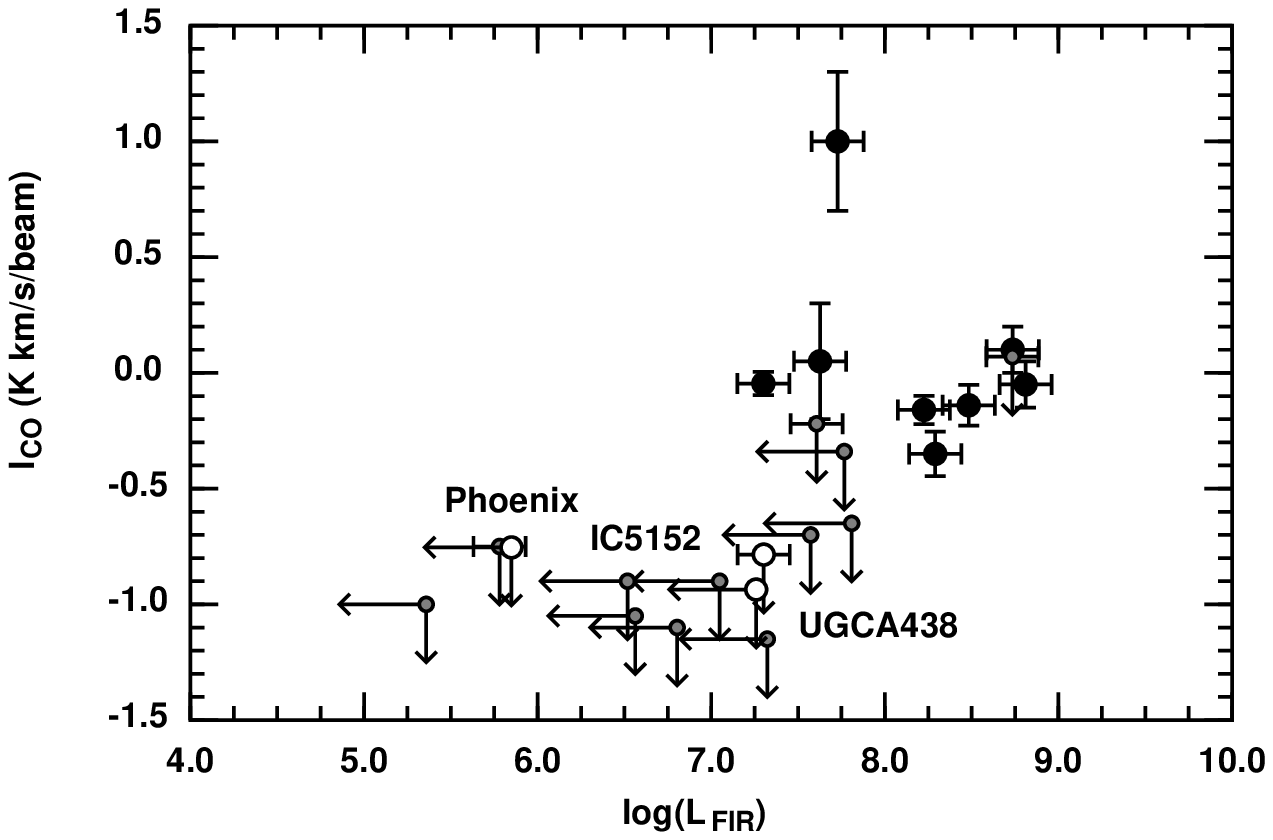"}
\caption{Left panel : the logarithm of the CO($1 \rightarrow 0$)
intensity, $I_{\rm CO}$, expressed in K~km~s$^{-1}$ per 55$''$ FWHM
beam, as in \citet{taylor1998}, versus the oxygen abundance,
$12+\log({\rm O/H})$. Right panel : the CO($1 \rightarrow 0$)
intensity, $I_{\rm CO}$ versus the logarithm of the far-infrared
luminosity, $L_{\rm FIR}$, calculated from the IRAS 60 and 100 micron
flux densities and expressed in solar bolometric luminosities. Black
data-points are dwarf galaxies detected by \citet{taylor1998}, grey
data-points are 4$\sigma$ upperlimits for the objects not detected by
\citet{taylor1998} and \citet{taylor2001}, white data-points are
4$\sigma$ upperlimits for the three galaxies presented in this
paper. For galaxies that were not detected by IRAS, we have assumed a
limiting 60 and 100 micron flux density of 1~Jy. Clearly, no dwarfs
with $12+\log({\rm O/H}) \la 7.9$ or $L_{\rm FIR} \la 7.5$ have been
detected in CO($1 \rightarrow 0$) emission so far. \label{COOH}}
\end{figure*}

The metallicity of the ionised gas in the star formation regions in
IC5152 has been determined using optical spectroscopy of the HII
regions, and ranges from $12 + \log({\rm O/H}) = 7.92 - 8.35$
\citep{webster1983,hidalgo2002,lee2003}. Phoenix and UGCA438 are not
presently forming stars, hence deriving a measurement for their
metallicity of the ISM is much more difficult. However in dIrrs, the
$B$-band luminosity is a good predictor of metallicity and we can use
the relation \citep{skillman2003}
\begin{equation}
  12 + \log({\rm O/H}) = (5.47 \pm 0.48) + (-0.147 \pm 0.029) M_B
\end{equation}
to estimate the nebular metallicity for these galaxies. In the case of
IC5152, this estimate gives $12+\log({\rm O/H}) = 8.00$, consistent
with the observations. Analogously, we find oxygen abundances
$12+\log({\rm O/H}) = 6.87$ and $7.32$ for Phoenix and UGCA438,
respectively. These values, as shown in Table~1, are broadly
consistent with the observed [Fe/H] abundances, assuming a solar
oxygen abundance of $12 + \log({\rm O/H})=8.7$ \citep{pla01}. In
Fig. \ref{COOH}, we add these three dwarf irregular galaxies to those
observed by \cite{taylor1998} and \cite{taylor2001}. We rescaled the
CO upper limits measured by us to a 55$''$ FWHM beam so as to match
the units in the figures presented in \citet{taylor1998} and
\citet{taylor2001}. The far-infrared luminosity is defined as $L_{\rm
FIR} = 3.65 \times 10^5\, D^2 f_{\rm FIR}\,L_\odot$. Here, $D$ is the
distance, expressed in Mpc, and $f_{\rm FIR} = 2.58\,f_{60} +
f_{100}$, with $f_{60}$ and $f_{100}$ the IRAS 60 and 100 micron flux
densities, respectively. For galaxies that were not detected by IRAS,
we have assumed a limiting 60 and 100 micron flux density of
1~Jy. Since the thermal emission of dust peaks between 50 and 100
micron, $L_{\rm FIR}$ is an indicator of the dust mass.

In a theoretical study of the formation of \Htwo\ at high redshift,
\citet{norman1997} showed that cold giant molecular clouds do not form
until the metallicity rises to $\sim 0.03 - 0.1$ Z$_\odot$, or
$12+\log({\rm O/H}) = 7.0-7.6$, at which point there is enough dust to
shield molecular clouds from the stellar UV radiation and metals are
sufficiently abundant to allow the gas to cool to a few tens of
Kelvin. Before that, star formation proceeds in a slow and
self-regulated way, as is also observed in nearby dwarf irregular
galaxies. As is clear from Fig. \ref{COOH}, no dwarfs with
$12+\log({\rm O/H}) \la 7.9$ or $\log(L_{\rm FIR}) \la 7.5$ have been
detected in CO so far \citep{taylor1998,leroy2005}. This seems to
indicate that the non-detections of CO in low-metallicity dIrrs
reflect a real lack of dust and hence cold molecular gas in these
systems.

An alternative explanation for the non-detections could be a large
increase in the CO-to-\Htwo\ conversion factor as the metal abundance
in the gas decreases \citep{leroy2005}.


\subsection{Phoenix : the dIrr to dSph transition}

The Phoenix dwarf is an interesting object. The H{\sc i} cloud
associated with the Phoenix dwarf is offset with respect to the
stellar distribution, probably due to ram pressure stripping by the
intergroup medium. Although some traces of molecular gas may still
remain within the galaxy's stellar body, cf. our tentative detection
of CO($1 \rightarrow 0$) emission at one position, it is unlikely that
a new star formation episode can start without a new inflow of neutral
gas. This transition-type dIrr/dSph galaxy might therefore evolve into
a dSph. No CO emission was found in the H{\sc i} cloud. This leaves
room for two interpretations. The lack of CO might indicate that ({\em i})
molecular clouds are not affected by ram pressure stripping in the
same way as the neutral gas, as shown by other observations
\citep{crowl2005}, or that ({\em ii}) simply no CO is present (like for most
low metallicity dwarfs). However, deeper CO observations or
observations using other molecular gas tracers are necessary to
confirm the evolutionary status of the Phoenix dwarf.

\section{Conclusions}
\label{sec_concl}

We have obtained very deep CO emission maps of three Local Group dwarf
irregular galaxies: IC5152, the Phoenix dwarf, and UGCA438. We did not
detect CO emission in these galaxies, with 4$\sigma$ upper limits of
$0.03-0.06$~K~km~s$^{-1}$ over a 6~km/s velocity width. Our results
are summarized in Table \ref{tab_results}. As is obvious from the
H{\sc i} maps of these galaxies, our maps cover both their stellar and
H{\sc i} distributions.

There could be two possible explanations for the lack of CO emission
in these galaxies. On the one hand, there could be a genuine lack of
molecular gas in these systems, in spite of the presence of large
amounts of neutral gas. On the other hand, the CO-to-\Htwo\ conversion
factor might become very large in these very low-metallicity dwarfs,
making CO a poor tracer of the molecular gas content. If this is
indeed the case, this has implications for galaxy formation scenarios
in the early universe, since molecules are an important coolant of the
ISM. In the absence of CO, star formation might not proceed as we
observe it today in our Galaxy. We find no correlation between the
composition of the interstellar medium and the star-formation
rate. All galaxies have H{\sc i} associated with them and Phoenix is
the only one with a tentative CO detection. However, only IC5152 is
actively forming stars. In the case of the Phoenix dwarf, we are most
likely witnessing the process of the transition of a dIrr to a dSph
galaxy, where the neutral gas has been stripped from the galaxy. There
might be a hint that some molecular gas is still present in the
galaxy, but this needs further confirmation.

\section*{Acknowledgements}

We wish to thank Bob Sault for granting us director telescope time. PB
and SDR acknowledge the Fund for Scientific Research Flanders (FWO)
for financial support. DM acknowledges financial support from the
Marie Curie MAGPOP Network.

\label{lastpage}

\end{document}